\documentclass[aps,preprint,prl,superscriptaddress,amsmath,amssymb,amsfonts]{revtex4-2}
\usepackage{graphicx}  
\usepackage{color}
\usepackage[colorlinks,bookmarks=true,citecolor=blue,linkcolor=blue,urlcolor=blue, breaklinks=true]{hyperref}
\usepackage{bm}
\usepackage{amsmath,amssymb}
\setcitestyle{super}

\def\q{{ {\bm q} }}
\def\s{{\sigma}}

\begin{document}

\title{Microscopic evidence for imaginary charge density wave in a kagome metal}

\author{S. Suetsugu}
\affiliation{Department of Applied Physics, The University of Tokyo, Tokyo 113-8656, Japan}
\affiliation{Department of Physics, Kyoto University, Kyoto 606-8502, Japan}
\author{F. Hori}
\affiliation{Department of Physics, Kyoto University, Kyoto 606-8502, Japan}
\affiliation{Department of Physics, Graduate School of Science, Tohoku University, 6-3 Aramaki-Aoba, Aoba-ku, Sendai, Miyagi 980-8578, Japan}
\author{M. Shibata}
\affiliation{Department of Physics, Kyoto University, Kyoto 606-8502, Japan}
\author{S. Kitagawa}
\affiliation{Department of Physics, Kyoto University, Kyoto 606-8502, Japan}
\author{K. Ishida}
\affiliation{Department of Physics, Kyoto University, Kyoto 606-8502, Japan}
\author{T. Asaba }
\affiliation{Department of Physics, Kyoto University, Kyoto 606-8502, Japan}
\affiliation{Department of Physics, University of Virginia, Charlottesville, VA. 22904, USA}
\author{S. Nakazawa}
\affiliation{Department of Physics, Nagoya University, Furo-cho, Nagoya 464-8602, Japan}
\author{Q. Li}
\affiliation{National Laboratory of Solid State Microstructures and Department of Physics, Nanjing University, Nanjing, China}
\author{H. -H. Wen}
\affiliation{National Laboratory of Solid State Microstructures and Department of Physics, Nanjing University, Nanjing, China}
\author{T. Shibauchi}
\affiliation{Department of Advanced Materials Science, University of Tokyo, Chiba, Japan}
\author{H. Kontani}
\affiliation{Department of Physics, Nagoya University, Furo-cho, Nagoya 464-8602, Japan}
\author{Y. Matsuda}
\affiliation{Department of Physics, Kyoto University, Kyoto 606-8502, Japan}
\affiliation{Los Alamos National Laboratory, Los Alamos, New Mexico 87545, USA}

\date{\today}





\maketitle
{\bf Dissipationless charge transport without any energy loss is one of the most fascinating phenomena in condensed matter physics. This extraordinary state manifests in two well-established systems: superconductors and quantum Hall systems. A proposed third category is associated with chiral loop current order, characterized by the spontaneous formation of microscopic electric current loops. The microscopic origin of these currents stems from imaginary hopping terms, conceptualized as an imaginary charge density wave (iCDW). Despite extensive investigations, its existence remains highly controversial. Here we report site-selective spectroscopic evidence for a pure iCDW in the kagome nonmagnetic metal CsV$_3$Sb$_5$. Nuclear quadrupole resonance spectra at out-of-plane $^{121}$Sb site sensitive to in-plane currents reveal anomalous broadening below $T^*\approx$120\,K, coinciding with the nematic transition well above the real charge density wave (CDW). Under magnetic fields, the spectra exhibit asymmetric lineshapes, demonstrating that this broadening purely originates from magnetic effects rather than from electric quadrupolar effects associated with CDW fluctuations. The observed lineshapes are quantitatively consistent with $\sim 1$\,mT local fields induced by chiral loop currents, indicating spontaneous time-reversal symmetry breaking. This microscopic identification of the long-sought pure iCDW establishes a novel form of quantum order, potentially revolutionizing our understanding of exotic electronic states in quantum materials.}

Dissipationless charge transport manifests in various quantum systems through distinct mechanisms. In superconductors \cite{tinkham2004introduction}, it emerges from spontaneous $U(1)$ gauge symmetry breaking, culminating in the formation of a coherent macroscopic quantum state. Quantum Hall systems \cite{PhysRevLett.45.494} exhibit dissipationless transport due to the topological protection of edge states, resulting in quantized Hall conductance intrinsically linked to the nontrivial topology of the system.
The chiral loop current order \cite{PhysRevB.37.3774,PhysRevLett.61.2015,PhysRevB.55.14554,varma2014pseudogap}, a manifestation of the third case of dissipationless current, arises from spontaneous symmetry breaking induced by quantum many-body effects in strongly correlated electron systems. This phenomenon has been postulated as a significant contributor to the enigmatic pseudogap phase observed in high-temperature superconductors, particularly in underdoped cuprates \cite{PhysRevB.37.3774,PhysRevB.55.14554,varma2014pseudogap,li2008unusual,PhysRevLett.111.047005}. Characterized by circulating currents within the CuO$_2$ planes, this ordered state offers a potential explanation for the partial gap in the electronic density of states above the superconducting transition temperature $T_c$.
Similar loop current behavior has been proposed in the context of a possible anapole state in the spin-orbit coupled Mott insulator Sr$_2$IrO$_4$ \cite{zhao2016evidence,jeong2017time,PhysRevX.11.011021}, whose low-energy electronic structure exhibits notable similarities to cuprates \cite{PhysRevLett.106.136402}. Despite extensive investigations, the presence of loop current order has remained highly controversial \cite{PhysRevLett.101.017001,PhysRevLett.106.097003,PhysRevB.96.214504}. While polarized neutron diffraction measurements have reported unconventional magnetic order in the pseudogap state of cuprates \cite{li2008unusual} and above the N\`{e}el temperature in iridates \cite{jeong2017time}, these results remain highly contentious \cite{PhysRevB.96.214504,zhang2018discovery,PhysRevB.101.195108}. Complementary muon spin relaxation ($\mu$SR) measurements on both cuprate and iridate systems indicate a possible critical slowing down of electronic fluctuations at relevant temperatures \cite{zhang2018discovery,PhysRevB.101.195108}. Collectively, these experimental observations suggest that loop currents, if present, manifest as a dynamic fluctuation phenomenon rather than a static order, with their detection critically dependent on the characteristic timescale of the experimental probe employed.

The kagome lattice, in which strong inherent geometrical frustration suppresses conventional orders such as spin density wave, serves as a fertile ground for investigating a diverse array of exotic quantum phases including the loop current order. Of particular interest are the kagome metals $A$V$_3$Sb$_5$ ($A$=K, Rb, Cs) \cite{PhysRevMaterials.3.094407,PhysRevLett.125.247002}, which have garnered significant attention due to their unique properties. These materials, whose crystal structure comprises two-dimensional (2D) V-Sb kagome layers (Figs. 1a and b), exhibit both CDW (or bond ordering) and superconducting instabilities. A focal point of research in these systems is the potential existence of a chiral loop current order or chiral flux phase \cite{PhysRevB.104.035142,PhysRevB.104.045122,PhysRevB.106.144504,PhysRevLett.127.217601,Tazai2023,Tazai-Morb,Shimura2023}, intimately connected to time-reversal-symmetry breaking. While some experimental evidence suggests the presence of chiral loop current orders near or well below the CDW transition temperature $T_{\rm CDW}$ \cite{yang2020giant,PhysRevB.104.L041103,jiang2021unconventional,guo2022switchable,mielke2022time,PhysRevResearch.4.023244,guguchia2023tunable,graham2024depth,xu2022three,hu2022time}, the nature and even the existence of this phenomenon remain subjects of intense debate \cite{PhysRevLett.131.016901,PhysRevB.105.045102,PhysRevB.110.195109,guo2024correlated}. The complexity of these systems is further compounded by the intricate coupling between the CDW and loop current, should the latter indeed be present.

Recent experimental investigations have revealed the existence of an odd-parity nematic order in kagome systems at temperatures significantly exceeding the CDW transition temperature, $T_{\rm CDW}$ \cite{asaba2024evidence}. This newly observed phase, characterized by broken rotational and time-reversal symmetry, has been tentatively associated with a potential loop current order (Fig.\,1c). However, the precise mechanism underlying this transition remains largely enigmatic and warrants further investigation.
The emergence of this nematic phase at temperatures substantially higher than $T_{\rm CDW}$ presents a unique opportunity to study symmetry-breaking phenomena in kagome systems in isolation from the confounding effects of the CDW formation. This temperature separation potentially facilitates a more rigorous elucidation of the fundamental physics governing these exotic orders, unencumbered by the intertwining interplay between CDW and chiral loop currents that typically complicates the phenomenon at lower temperatures.
In this context, it is noteworthy that the chiral loop current that occurs well above $T_{\rm CDW}$ arises from purely imaginary hopping processes (Fig.\,1d). Consequently, the chiral loop current can be conceptualized as a manifestation of a purely imaginary CDW (iCDW) \cite{PhysRevB.62.4880,PhysRevB.83.020505,PhysRevB.91.201105}, which is fundamentally distinct from conventional real CDW (rCDW) (Figs.\,1e and f). This distinction provides a novel framework for understanding the diverse symmetry-breaking mechanisms in kagome systems and may offer new insights into the rich phase diagram of these materials.

Nuclear quadrupole resonance (NQR) and nuclear magnetic resonance (NMR) spectroscopy serve as a powerful and the most direct microscopic probe for investigating the chiral loop current order. The spontaneous loop current  generate local internal magnetic fields, which can potentially be detected through their effects on spectral line shifts in NQR/NMR experiments. In this study, by performing $^{121}$Sb NQR/NMR experiments, we provide the microscopic evidence for the time reversal symmetry breaking well above $T_{\rm CDW}$. The present results demonstrate the emergence of the long-sought pure imaginary CDW phase in the kagome metal CsV$_3$Sb$_5$.

The kagome metal CsV$_3$Sb$_5$ has two distinct sites of Sb \cite{PhysRevMaterials.3.094407}. Sb1 is positioned at the center of the V hexagon, and Sb2 is located above and below the V triangle (Figs. 1a and b). The NQR frequency is given by $\nu_q = \frac{3e^2qQ}{2I(I-1)h}$, where $eq$ represents the electric field gradient (EFG), $Q$ is the nuclear quadrupole moment, $I$ is the nuclear spin, and $h$ is the Planck constant. Extended Data Figure\,1 shows the zero-field NQR spectra of $^{121}$Sb ($I=5/2$) corresponding to $\pm 1/2 \leftrightarrow \pm 3/2$ transitions at 100\,K. The spectral intensity ratio reflects the atomic ratio of Sb1:Sb2 = 1:4. In this study, we mainly discuss the $^{121}$Sb2 spectra, which are stronger and sharper than those of $^{121}$Sb1 (see Supplementary Information for $^{121}$Sb1 spectra).

Remarkably, the $^{121}$Sb2 NMR spectra exhibit progressive linewidth broadening upon cooling from elevated temperatures down to the CDW transition temperature $T_\mathrm{CDW} = 94$\,K (Fig.\,2a). To quantify this broadening systematically, we analyzed the temperature dependence of the full-width at half-maximum (FWHM) $\delta\nu_q$ as shown in Fig.\,2b (black circles). Upon decreasing temperature, $\delta\nu_q$ initially increases gradually, followed by a pronounced enhancement below $T^* \sim120$\,K, at which a kink-like anomaly might be present. This characteristic temperature is closely related to the nematic transition temperature $T_\mathrm{nem}$ (inset of Fig.\,2b), at which spontaneous $C_6 \rightarrow C_2$ symmetry breaking occurs, as previously established through magnetic torque measurements \cite{asaba2024evidence}. The nematic transition temperature $T_\mathrm{nem}$ decreases with decreasing field, and a linear extrapolation to zero field yields a temperature very close to $T^*$. For comparison, we also plot the amplitude of two-fold oscillations of magnetic torque $\tau_{2\phi} / V$ measured in an in-plane field of 7\,T, where the nematic transition occurs at $T_\mathrm{nem}$(7\,T)=130\,K. While $\tau_{2\phi} / V$ remains zero above the nematic transition, the spectral linewidth begins to broaden already above $T^*$. Moreover, unlike the clear kink observed in $\tau_{2\phi} / V$ at $T_\mathrm{nem}$, the linewidth exhibits a broadened enhancement rather than a sharp anomaly. As discussed below, these contrasting behaviors reflect the presence of strong thermal fluctuations above the thermodynamic transition temperature.

The anomalous linewidth broadening provides compelling evidence for an additional contribution to the inhomogeneous linewidth due to magnetic or electric quadrupole interactions. In the latter case, the broadening likely originates from CDW fluctuations. However, this scenario can be questioned for the following reasons. Below $T_\mathrm{CDW} = 94$\,K, the normal state peak is suppressed and multiple peaks emerge due to CDW modulation \cite{luo2022possible}. As shown in Extended Data Fig.\,2, both normal-state (red arrows) and CDW-state peaks (blue arrows) are present at 93 and 94\,K, indicating the CDW transition is first-order. While previous studies have attributed a Curie-Weiss-like temperature dependence of $\delta\nu_q$ above $T_\mathrm{CDW}$ to a short-range CDW order arising from CDW fluctuations \cite{feng2023commensurate}, the first-order character of the CDW transition suggests that CDW fluctuations are unlikely to be strongly developed. Moreover, it should be noted that a microscopic theoretical framework that quantitatively relates CDW fluctuations to the linewidth $\delta\nu_q$ is still lacking. Even if the CDW fluctuations were present, our field-dependent experiments discussed below demonstrate the magnetic origin of the linewidth broadening.

To disentangle magnetic and quadrupolar contributions to the anomalous broadening, we performed $^{121}$Sb2 NMR experiments under magnetic fields $\bm{H}$ applied along the crystallographic $c$ axis. Since the principal axis of EFG at the Sb2 site is also aligned along the $c$ axis \cite{luo2022possible}, the applied field $\bm{H}||c$ just split the degenerate resonant frequency $\nu_q$ into $f(H_c) = \nu_q \mp \gamma\mu_0(H_c + \Delta H)$, corresponding to the $\pm1/2 \leftrightarrow \pm3/2$ transitions (Fig.\,3a). Here $\gamma=10.189$\,MHz/T is the gyromagnetic ratio of $^{121}$Sb nuclei \cite{harris2008further}, and $\Delta H$ represents the hyperfine field at the Sb2 site. Figure 3d shows the $^{121}$Sb2 NMR spectra of the $+1/2 \leftrightarrow +3/2$ transition measured at $\mu_0H_c \sim 0.95$\,T for temperatures above $T_\mathrm{CDW}$. The spectra exhibit asymmetric lineshapes, with tails extending toward higher frequencies. While the linewidth increase with decreasing $T$ is similar to that observed in the NQR data (Fig.\,2a), this broadening is distinctly asymmetric, with the high-frequency tail becoming more pronounced at lower temperatures. In contrast, the $-1/2 \leftrightarrow -3/2$ spectra (Fig.\,3e) display tails extending toward lower frequencies, opposite to the trend observed in the $+1/2 \leftrightarrow +3/2$ spectra. To quantitatively characterize this behavior, Fig.\,3f plots the temperature dependence of the difference between the peak frequency and the spectral centroid, $\nu_\mathrm{peak}-\nu_\mathrm{c}$ (see Methods). Below $T^{\ast}$, $\nu_\mathrm{peak}-\nu_\mathrm{c}$ increases for the $-1/2 \leftrightarrow -3/2$ transition, whereas it decreases for the $+1/2 \leftrightarrow +3/2$ transition, directly reflecting the antisymmetric character observed in Figs.\,3d and 3e.

This antisymmetric behavior between the $\pm 1/2 \leftrightarrow \pm 3/2$ transitions arises from the opposite sign of the hyperfine fields shift $\mp \gamma\mu_0\Delta H$, depending on the nuclear spin state (Fig.\,3b). In zero field, $\nu_\mathrm{peak}-\nu_\mathrm{c}$ for the $\pm 1/2 \leftrightarrow \pm 3/2$ transition remains zero, indicating a symmetric lineshape resulting from the overlap of antisymmetric spectra. Application of a magnetic field lifts this degeneracy, revealing the intrinsic antisymmetric character. The emergence of such antisymmetric lineshapes provides direct microscopic evidence for local internal magnetic fields, demonstrating time reversal symmetry breaking above $T_\mathrm{CDW}$. By contrast, the quadrupole shift $\nu_q$ is independent of the sign of the nuclear spin, yielding symmetric spectra between these transitions (Fig.\,3c). Indeed, the CDW state peak appears at lower frequencies relative to the normal state peak in both transitions (Extended Data Fig.\,3), reflecting a uniform quadrupole shift. This behavior confirms that quadrupole interactions contribute equally to both transitions and cannot account for the observed antisymmetric lineshapes. This conclusion remains valid even for a finite EFG asymmetry parameter $\eta$, that cannot produce the observed antisymmetric lineshapes (see Supplementary Information). These results further corroborate the magnetic origin of the linewidth broadening.

Having established the magnetic origin of the anomalous broadening above $T_\mathrm{CDW}$, we turn to the source of the emergent hyperfine fields $\Delta H$. In CsV$_3$Sb$_5$ with no magnetic ions, the most plausible explanation is the emergence of an iCDW state, in which loop currents generate internal magnetic fields. Theoretically, two distinct current configurations of triple-$\bm{q}$ and single-$\bm{q}$ states have been widely considered \cite{PhysRevB.104.035142,PhysRevB.104.045122,Tazai2023,Tazai-Morb,Shimura2023,asaba2024evidence}. Under a strong external field $\bm{H}||\bm{c}$, the $^{121}$Sb2 resonant frequency is primarily sensitive to the $c$-axis component of the local fields, as captured by first-order perturbation theory. We therefore restrict our analysis to this out-of-plane component. In the triple-$\bm{q}$ loop current order, the Sb2 sites experience an asymmetric distribution of hyperfine fields, with $\Delta H = +h_\mathrm{loc}$ and $-\frac{1}{3}h'_\mathrm{loc}$ in a 1:3 ratio (Fig.\,4a), consistent with the observed asymmetric lineshape. Here $h'_\mathrm{loc}$ is approximately $h_\mathrm{loc}$ according to Ref.\,\citenum{Shimura2023}. In contrast, the single-$\bm{q}$ state produces symmetric local fields of $+h_\mathrm{loc}$ and $-h_\mathrm{loc}$ with equal weight (Fig.\,4b), which would yield a symmetric spectrum. Therefore, the pronounced spectral asymmetry suggests the realization of a triple-$\bm{q}$ loop current ordered state.

To quantify the emergent hyperfine fields, we fitted the asymmetric spectra using a model based on the triple-$\bm{q}$ loop current ordered state. The spectral profile is expressed as a weighted sum of two components, corresponding to $\Delta H = +h_\mathrm{loc}$ and $-\frac{1}{3}h_\mathrm{loc}$ in a 1:3 ratio (see Methods). As shown in Fig.\,4c, this model offers excellent fits to the experimental data, with the best fit obtained for $h_\mathrm{loc} = 30$\,Oe. Moreover, as shown in Extended Data Figs.\,4 and 5, this model well reproduces the experimental data across a broad range of temperatures and external fields. The present results indicate the observed spectral broadening is quantitatively consistent with internal magnetic fields induced by triple-$\bm{q}$ loop currents.

We emphasize that the time reversal symmetry in CsV$_3$Sb$_5$ is spontaneously broken in zero field, rather than being induced by external fields.
The anomalous linewidth broadening in zero field, which appears symmetric due to the degeneracy of the $\pm 1/2 \leftrightarrow \pm 3/2$ transitions, manifests as the pronounced asymmetric broadening when an external magnetic field is applied. This behavior indicates that the same internal magnetic fields are responsible for both the symmetric linewidth broadening in zero field and the asymmetric broadening in external fields. This conclusion is further supported by  the following reasons. The temperature dependence of $h_\mathrm{loc}$ at $\mu_0H \sim$\,0.5, 0.95, and 1.5\,T exhibits a sharp increase below $T^* \approx 120$\,K (Fig.\,4d and inset), closely matching the onset of the linewidth broadening observed in the zero-field NQR measurements. As shown in Fig.\,4e, $h_\mathrm{loc}$ increases nearly linearly with external magnetic field $H$, indicating that the current order parameter grows with $H$. Remarkably, extrapolation to zero field yields a finite $h_\mathrm{loc} = 8 \pm 2$\,Oe, which provides compelling evidence for spontaneous time reversal symmetry breaking in CsV$_3$Sb$_5$. This extrapolated value is in excellent agreement with theoretical calculations, that predict $h_\mathrm{loc} \approx 8$\,Oe for the triple-$\bm{q}$ state in CsV$_3$Sb$_5$ (see Extended Data Fig.\,6a). While the single-$\bm{q}$ configuration cannot be excluded based on the symmetric zero-field lineshape, the zero-field NQR spectra are well reproduced by the triple-$\bm{q}$ state with the $h_\mathrm{loc} \approx 8$\,Oe (see Extended Data Fig.\,7 and Methods).

Previous muon spin rotation ($\mathrm{\mu}$SR) experiments have reported time reversal symmetry breaking only below $T_\mathrm{CDW}$ in $A$V$_3$Sb$_5$ \cite{mielke2022time,PhysRevResearch.4.023244,guguchia2023tunable}. Furthermore, recent depth-dependent $\mathrm{\mu}$SR studies have suggested that time-reversal symmetry is broken above $T_\mathrm{CDW}$ only near the surface in RbV$_3$Sb$_5$ \cite{graham2024depth}. In contrast, the present bulk-sensitive NQR/NMR experiments reveal that the local fields emerge in the bulk above $T_\mathrm{CDW}$. These differences are likely due to the position at which the local magnetic fields are detected. While the present NMR experiments detect the local fields at the Sb2 site, $\mathrm{\mu}$SR measurements probe local fields at the muon stopping site. It is possible that the local field at the muon site is significantly weaker than that at the Sb2 site. While previous Kerr effect experiments have reported conflicting results regarding the time reversal symmetry breaking below $T_\mathrm{CDW}$ \cite{xu2022three,hu2022time,PhysRevLett.131.016901}, none of them have reported the time reversal symmetry breaking above $T_\mathrm{CDW}$. We point out that, in both triple-$\bm{q}$ and single-$\bm{q}$ states, opposing magnetic fields cancel out within the optical spot size used in the Kerr effect measurements, making the detection of the local fields extremely challenging.

We next examine the effect of thermal fluctuations in the NQR/NMR spectra. As shown in Fig.\,2b, the zero-field linewidth broadening (black circles) begins well above $T^* \approx 120$\,K. This behavior differs markedly from the twofold in-plane oscillations in magnetic torque measurements \cite{asaba2024evidence}, which remain zero above the critical temperature $T^*$. Above the transition temperature into the iCDW phase, thermal fluctuations of the current order parameter generate both right- and left-handed local loop currents. Although these fluctuating currents cancel out in macroscopic probes such as magnetic torque, they give rise to local magnetic fields that are readily detected in microscopic NQR/NMR experiments. Indeed, theoretical calculations that incorporate thermal fluctuations show that the fluctuating order parameter yields finite local fields even above the transition temperature (see Methods and Extended Data Fig.\,6b). The observed linewidth broadening is well captured by these calculations (blue dashed line in Fig.\,2b) using $T^* = 120$\,K determined by the zero-field extrapolation of the magnetic torque data \cite{asaba2024evidence}. The best fit is obtained for a dimensionless parameter $A = 1.14$, indicating strong thermal fluctuations. Within this framework, the finite $h_\mathrm{loc}$ values even at 160\,K (Fig.\,4d) are naturally explained by thermal fluctuations of the iCDW order.

Previous studies have reported signatures of time-reversal symmetry breaking in CsV$_3$Sb$_5$, but the onset temperature has remained highly controversial \cite{yang2020giant,PhysRevB.104.L041103,jiang2021unconventional,guo2022switchable,mielke2022time,PhysRevResearch.4.023244,guguchia2023tunable,graham2024depth,xu2022three,hu2022time}. The present $^{121}$Sb NQR/NMR measurements provide definitive microscopic evidence that time-reversal symmetry breaks spontaneously at $T^*$, significantly above the CDW transition temperature $T_\mathrm{CDW}$. The asymmetric lineshapes observed in the $^{121}$Sb NMR spectra point to a pure imaginary CDW state with triple-$\bm{q}$ loop currents, in which emergent local magnetic fields strengthen with decreasing temperature and increasing external field. These observations suggest that CsV$_3$Sb$_5$ first enters an imaginary CDW phase at $T^*$, characterized by modulation solely in the imaginary part of the hopping term, $\delta t = i\eta$. Upon further cooling below $T_\mathrm{CDW}$, a real component $\phi$ develops as well, leading to a complex modulation $\delta t = \phi + i\eta$ in which real CDW and loop current orders coexist. These findings offer crucial insights into the long-sought imaginary CDW order, establishing a foundation for exploring exotic electronic states in quantum materials.

\bibliography{ref.bib}

\noindent
{\bf METHODS}\\
\noindent
{\bf Single crystal growth}\\
High-quality single crystals of CsV$_3$Sb$_5$ were grown by a self-flux method with a Cs-Sb binary eutectic mixture as the flux \cite{xiang2021twofold}. High purity elements of Cs bulk, V piece, and Sb shot were used and prepared in the molar ratio of Cs:V:Sb = 7:3:14. Then the mixture was loaded into an aluminum crucible and sealed in an evacuated quartz tube, heated slowly to 1000\,$^\circ$C and stayed for 24\,h. It was cooled down to 200\,$^\circ$C in 260\,h and subsequently down to room temperature with the power of the furnace switched off. Shiny crystals with an apparent hexagonal edge were obtained after the flux was removed.

\noindent
{\bf NQR/NMR experiments}\\
$^{121}$Sb NQR/NMR measurements were conducted using a conventional spin-echo technique. For NMR experiments, magnetic fields were applied parallel to the crystallographic $c$-axis, which aligned with the principal axis of the EFG at the Sb2 site. A rotator was employed to ensure precise alignment of the fields along the $c$ axis. The NQR/NMR spectra were acquired through a combination of Fourier transformation and frequency sweep. The spectral linewidth $\delta\nu_q$, defined as the full-width at half-maximum (FWHM), was determined by fitting the spectra with a Voigt function.

Since it has been reported that an asymmetry parameter of the EFG $\eta$ is zero above $T_\mathrm{CDW}$ \cite{mu2021s}, we focus on the case of $\eta = 0$ in the main text. The effect of finite $\eta$ on the NMR spectra is discussed in detail in the Supplementary Information. For $\eta = 0$, the nuclear spin Hamiltonian $\mathcal{H}$ in the presence of an external magnetic field $\bm{H}$ is given by
\begin{align}
\mathcal{H} &= \mathcal{H}_Q + \mathcal{H}_Z \\
\mathcal{H}_Q &= \frac{h\nu_{q}}{6}(3I^2_z - I^2) \\
\mathcal{H}_Z &= -\gamma\hbar\mu_0(H + \Delta H) \nonumber\\
 & \times (I_z\cos\theta + \frac{I_+e^{-i\phi}+I_-e^{i\phi}}{2}\sin\theta),
\label{eqn:Hamiltonian}
\end{align}
where $\mathcal{H}_Q$ and $\mathcal{H}_Z$ are the electric quadrupole and Zeeman interaction terms, respectively. Here $\Delta H$ represents the hyperfine field at the nuclear site, $\theta$ and $\phi$ are the polar and azimuthal angles of the external field relative to the principal axis of EFG. Since the EFG principal axis at Sb2 is along the $c$ axis, the applied field $\bm{H}||c$ just split the degenerate resonant frequency $\nu_q$ of the $\pm 1/2 \leftrightarrow \pm 3/2$ transitions into $f(H_c) = \nu_q \mp \gamma\mu_0(H_c + \Delta H)$, corresponding to $+1/2 \leftrightarrow +3/2$ and $-1/2 \leftrightarrow -3/2$ transitions (Fig.\,3a). Therefore, hyperfine fields shift the resonant frequency with opposite signs depending on the nuclear spin state, leading to antisymmetric linewidth broadening between these transitions (Fig.\,3b). By contrast, the frequency shift originating from quadrupole interactions is independent of the sign of the nuclear spin, yielding symmetric broadening (Fig.\,3c). This distinction allows a comparative analysis of the lineshapes to reveal the presence and nature of hyperfine fields at the Sb2 sites. To quantitatively characterize the asymmetry of the spectral line shape, we analyzed the spectral centroid. The spectral centroid $\nu_\mathrm{c}$ was calculated as
\begin{equation}
\nu_\mathrm{c}
=
\frac{\int \nu\, I(\nu)\, d\nu}{\int I(\nu)\, d\nu},
\end{equation}
where $I(\nu)$ is the spectral intensity as a function of frequency $\nu$. The asymmetry was then quantified by the difference between the peak frequency and the spectral centroid, $\nu_\mathrm{peak}-\nu_\mathrm{c}$.

To quantify the internal local fields induced by emergent loop currents, we modeled the spectra assuming a triple-$\bm{q}$ configuration. In this state, the Sb2 sites experience hyperfine fields of $\Delta H = +h_\mathrm{loc}$ and $-\frac{1}{3}h_\mathrm{loc}$ in a 1:3 ratio. The total spectral intensity is given by $I_\mathrm{total}(f, \Delta) = \frac{A}{3}V(f + \Delta) + AV(f - \frac{1}{3}\Delta)$, where $A$ is the normalization factor, $V(f)$ is the Voigt function representing the intrinsic lineshape of Sb2, and $\Delta = \gamma \mu_0h_\mathrm{loc}$ is the frequency shift due to the local fields $h_\mathrm{loc}$. As shown in Fig.\,4c and Extended Data Figs.\,4 and 5, this model provides excellent fits to the experimental data. While the triple-$\bm{q}$ state is expected to host two types of domains with opposite local field directions, the observed asymmetric lineshapes suggest a field-induced alignment of the loop current domains (see Supplementary Information).

In zero field, the $+1/2 \leftrightarrow +3/2$ and $-1/2 \leftrightarrow -3/2$ transitions overlaps, and the total spectral intensity becomes
\begin{align}
I_\mathrm{total}(f, \Delta) &= \frac{A}{3}V(f + \Delta) + AV(f - \frac{1}{3}\Delta) \nonumber\\
& + \frac{A}{3}V(f - \Delta) + AV(f + \frac{1}{3}\Delta).
\label{eqn:nqr-spectra}
\end{align}
As shown in the Extended Data Fig.\,7, this spectral form yields an excellent fit to the zero-field data using the extrapolated zero-field value $h_\mathrm{loc} = 8$\,Oe obtained from the analysis on the field dependence.

\noindent
{\bf Numerical calculation of the internal field in the iCDW phase}\\
In kagome metals, $2\times2$ iCDW order breaks the time-reversal symmetry, leading to the emergence of the chiral motion of electrons.
The iCDW order is schematically shown in Extended Data Fig.\,8a. It is given by the pure imaginary modulation of the nearest-neighbor V sites $(i,j)$; $\delta t_{i,j}=i\eta$ or $-i\eta$ ($\eta$ is a real constant).
Note that the loop current order is odd-parity to satisfy the Hermite condition: $\delta t_{i,j}=-\delta t_{j,i}$.
Such imaginary $\delta t_{i,j}$ cannot be derived from conventional Peierls transition originating from the electron-lattice interaction.
However, the rCDW fluctuations mediate the $p$-wave electron-hole pairing condensation, which is nothing but the iCDW order parameter \cite{Tazai2023}.
The total hopping integral from site $j$ to $i$ is $t_{i,j}\equiv t_{i,j}^0+\delta t_{i,j}$, where $t_{i,j}^0$ is the original real hopping integral.
The band structure of the kagome metal is folded into the reduced Brillouin zone by the $2\times2$ iCDW, which transfers the three van Hove singularities (vHSs) at the M points to the $\mathrm{\Gamma}$ point. The iCDW order reconstructs the bands around $\mathrm{\Gamma}$, lifting the threefold degeneracy of the vHS. The resulting band dispersion for $\eta = -0.01$\,eV is shown in Extended Data Fig.\,8b.

Due to the broken time-reversal symmetry,
the iCDW order gives rise to dissipationless charge current from site $j$ to site $i$, $J_{i,j}$.
Importantly, $J_{i,j}$ is gauge invariant irrespective of the fact that $t_{i,j}$ depends on the choice of the gauge.
The inter-bond current is given as
$J_{i,j} = \langle j_{i,j} \rangle = it_{i,j} \langle c_{i,\s}^\dagger c_{j,\s} \rangle- \{ i \leftrightarrow j \}$.
Here, we derive $J_{i,j}$ in the $2\times2$ unit cell by following Ref. \citenum{Shimura2023}.
Next, we obtain the local magnetic field using the Biot-Savart law by summing the contributions from $J_{i,j}$ at sufficiently distant sites. 

In Extended Data Fig.\,6a, we present the numerical results of the local magnetic field at a Sb site with $+h_{\rm loc}$ (see Fig.\,4a).
Here, we set $\eta=0.01$\,eV because the relation $T^*\sim\eta$ is naturally expected theoretically.
($+h_{\rm loc}$ is approximately proportional to $\eta^1$.)
The obtained $h_{\rm loc}$ is of order 10\,Oe for $n\sim3$.
Note that the magnetic filed at other three Sb2 sites is approximately $-h_{\rm loc}/3$, while the cancellation of total magnetic field is imperfect.
Therefore, the uniform orbital magnetization remains finite in the triple-$\bm{q}$ iCDW state \cite{Tazai-Morb}.

\noindent
{\bf Effect of thermal fluctuations on the linewidth broadening}\\
To analyze the NQR/NMR linewidth broadening caused by thermal fluctuations of the loop current order parameter $\eta_{\bm{q}}$ at wavevector $\bm{q}$, we employ a simplified Ginzburg-Landau (GL) framework. The temperature dependence of the fluctuation-induced broadening is estimated from the mean square fluctuation $\langle \eta^2 \rangle$, as described below.

For $T > T^*$, the mean value of the order parameter vanishes, $\eta_0 = 0$.
Then, the GL free energy is given by
\begin{equation}
F[\eta_q] = a(1 + \xi_0^2 (\bm{q} - \bm{q}_m)^2)\eta_{\bm{q}}^2 + c\eta_{\bm{q}}^4,
\end{equation}
where $\xi_0 = |1 - T/T^*|^{-1/2}$ is the correlation length.
The second-order coefficient is $a = \alpha(T/T^* - 1)$ with $\alpha / N(0)$ = 0.01--0.1 \cite{Tazai-PRB2023}, where $N(0)$ is the density of states of conduction electrons. Here $\bm{q}_m$ ($m=1, 2, 3$) is the wavevector of the $m$-th iCDW. In the following, we replace $\bm{q}-\bm{q}_m$ with $\bm{q}$ to simplify the notation. In this regime, the mean square fluctuation is obtained as
\begin{eqnarray}
\langle \eta^2 \rangle
&\approx& \frac{1}{N} \sum_\q^{|\q|<q_c} \langle\eta_\q^2\rangle
\nonumber \\
&=& \frac1{16\pi}\frac1{a\beta\xi^2} \ln(\xi^2+1) \equiv H^2_a,
\label{eqn:J1}
\end{eqnarray}
where $\displaystyle \frac1N \sum_\q^{|\q|<q_c}\cdots =\int_0^1\frac{qdq}{2(2\pi)^2}\cdots$ and $\xi^2 = (\xi_0^{-2} + L^{-2})^{-1}$. 
We have introduced a cutoff $q_c$ of order 1 in the $q$-integration. Numerical results are insensitive to the choice of $q_c$ because the contribution from $|\q| \gg \xi_0^{-1}$ is very small. Hereafter, we set $\beta = 1/T^*$ for simplicity.

For $T < T^*$, the order parameter develops a finite mean value $\eta_0 \sim T^* \sqrt{1 - T/T^*}$, and the GL free energy becomes
\begin{eqnarray}
F[{\tilde\eta}_\q]= a'(1+\xi'^2(\q-\q_m)^2){\tilde\eta}_\q^2,
\end{eqnarray}
with $a' = -2a$, $\xi'^2_0 = \xi_0^2/2$, and $\tilde{\eta}_q = \eta_q - \eta_0$. The fluctuation amplitude is similarly approximated as
\begin{eqnarray}
\langle {\tilde\eta}^2 \rangle \approx
\frac1N \sum_\q^{|\q|<1} \langle{\tilde\eta}_\q^2\rangle \equiv H^2_{a'}.
\end{eqnarray}

To capture the linewidth broadening across the entire temperature range, we define a unified expression approximately valid across the transition as
\begin{eqnarray}
(\delta\nu/\nu)^2 &\propto& \langle {\eta}^2 \rangle \approx {\tilde H}^2_{a'} ,
\label{eqn:J2}
\\
{\tilde H}^2_{a'} &\approx& [(\eta_0+H_{a'})^2+(\eta_0-H_{a'})^2]/2.
\label{eqn:H}
\end{eqnarray}
The numerically calculated $\langle \eta^2 \rangle$ is plotted in Extended Data Fig. 6b. Here we set $a = a'$ for simplicity and use $L^2 = 30$ that represents a smooth cutoff to satisfy $\xi^2 \lesssim L^2$. The strength of thermal fluctuations is governed by a dimensionless parameter $A \equiv 1/\alpha T^*$. Note that the parameter $A$ is typically in the range 0.3--1 for the correlation-driven off-site bond orders (rCDW and iCDW), such as the electronic nematic state of FeSe \cite{Tazai-PRB2023}. To estimate the total linewidth observed in experiments, we add a constant offset to $\langle \eta^2 \rangle$ representing the intrinsic linewidth $\delta\nu_0$ of the Sb2 site using the error propagation formula, $(\delta\nu_q / \nu_q)^2 = {\tilde H}_a^2 + (\delta\nu_0 / \nu_q)^2$. As shown by the blue dashed line in Fig.\,2b, this yields an excellent fit to the experimental data using $T^* = 120$\,K determined from the magnetic torque experiments \cite{asaba2024evidence}. Here the best fit is obtained for $A = 1.14$, indicating strong thermal fluctuations.

\noindent
{\bf Data and materials availability}\\
Other data that support the findings of this study are available from the corresponding author upon request. Source data are provided with this paper for the main figures and Extended Data figures.

\noindent
{\bf Acknowledgments}\\
We thank A. Furusaki and Guo-qing Zheng for insightful discussions. This work is supported by Grants-in-Aid for Scientific Research (KAKENHI) (Nos. JP23K13060, JP23H00089, JP25K00958, JP24K00568, and JP22H00105), and Transformative Research Areas A “Extreme Universe” (No. 24H00965) and “Correlated Design Science” (No. JP25H01248) from the Japan Society for the Promotion of Science, and CREST (JPMJCR19T5) and PRESTO (No. JPMJPR2252; T.A.) from the Japan Science and Technology (JST). 

\noindent
{\bf Author contributions}\\
S.S, T.A., and Y.M. conceived the projects. Q.L. and H.-H.W. synthesized the high-quality single crystals. S.S., F.H., M.S., S.K., and K.I. conducted NQR and NMR experiments. H.K. performed theoretical calculations. All authors analyzed the data and discussed the results. S.S., T.S., H.K., and Y.M. prepared the manuscript with inputs from all authors.

\noindent
{\bf Competing interests}\\
The authors declare no competing interests.

\clearpage

\begin{figure}
	\includegraphics[clip,width=8.5cm]{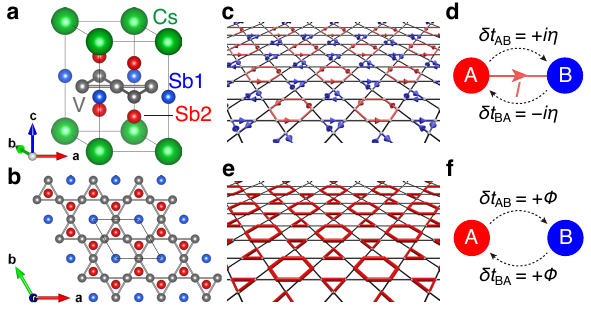}
	\caption{{\bf Crystal structure, loop current, and charge density wave orders.} {\bf a}, {\bf b}, Crystal structure of the kagome metal CsV$_3$Sb$_5$, which hosts two inequivalent Sb sites. While Sb1 resides at the center of the V hexagon, Sb2 is located above and below the V triangle. {\bf c}, Schematic of the triple-$\bm{q}$ loop current ordered state of a kagome metal. Spontaneous loop currents circulate along red hexagons and blue triangles. {\bf d}, Imaginary charge density wave (iCDW) state. Modulation in the imaginary component of the hopping amplitude $\delta t = i\eta$ gives rise to spontaneous currents $I$. {\bf e}, Triple-$\bm{q}$ bond ordered (Tri-hexagonal CDW) state of a kagome metal. {\bf f}, Real charge density wave (rCDW) state. Modulation in the real part of the hopping amplitude leads to bond modulation without current.
	\label{fig:crystal}
	}
\end{figure}

\begin{figure}
	\includegraphics[clip,width=8.5cm]{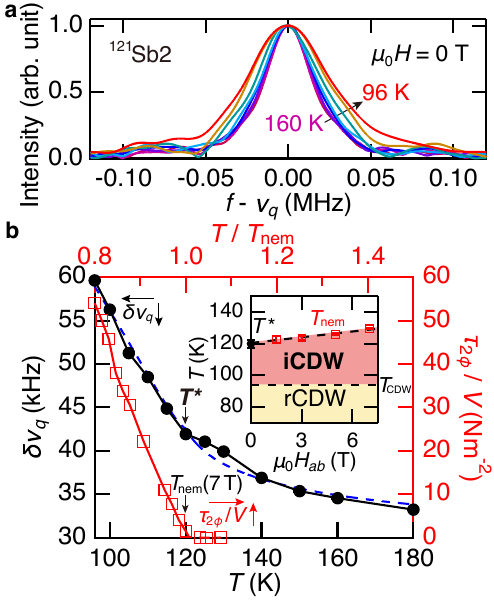}
	\caption{{\bf Temperature dependence of $^{121}$Sb2 NQR spectra.} {\bf a}, $^{121}$Sb2 NQR spectra measured at various temperatures down to 96\,K. The normalized intensity is plotted as a function of $f - \nu_q$ to highlight the anomalous linewidth broadening that emerges well above the CDW transition temperature $T_\mathrm{CDW} = 94$\,K. {\bf b}, Temperature dependence of the linewidth $\delta\nu_q$. A pronounced increase is observed below the characteristic temperature $T^\ast \approx 120$\,K. The error bars estimated by the standard errors of the fitting parameters are smaller than the symbol size. For comparison, the amplitude of two-fold oscillations of magnetic torque measured in an in-plane field of 7\,T is also shown (red squares), plotted on the top axis as a function of $T/T_\mathrm{nem}$ with $T_\mathrm{nem}(7\,\mathrm{T}) = 130$\,K. The blue dashed line shows the theoretically calculated root-mean-square fluctuation of the current order parameter $\sqrt{\langle \eta^2 \rangle}$ incorporating thermal fluctuations (see Methods). To enable the comparison with the experimental data, a constant offset representing the intrinsic $^{121}$Sb2 linewidth is added to $\langle \eta^2 \rangle$. The inset displays the field dependence of the nematic transition temperature $T_\mathrm{nem}$ (red squares), determined from twofold in-plane oscillations in magnetic torque measurements \cite{asaba2024evidence}. The namatic transition temperature gradually decreases with decreasing $H$. A linear extrapolation of $T_\mathrm{nem}$ to zero field yields a temperature very close to $T^*$ (black circle). The error bar of $T^*$ represents half of the temperature step of the NQR measurements.
	\label{fig:NQR} 
	}
\end{figure}

\begin{figure*}
	\includegraphics[clip,width=16cm]{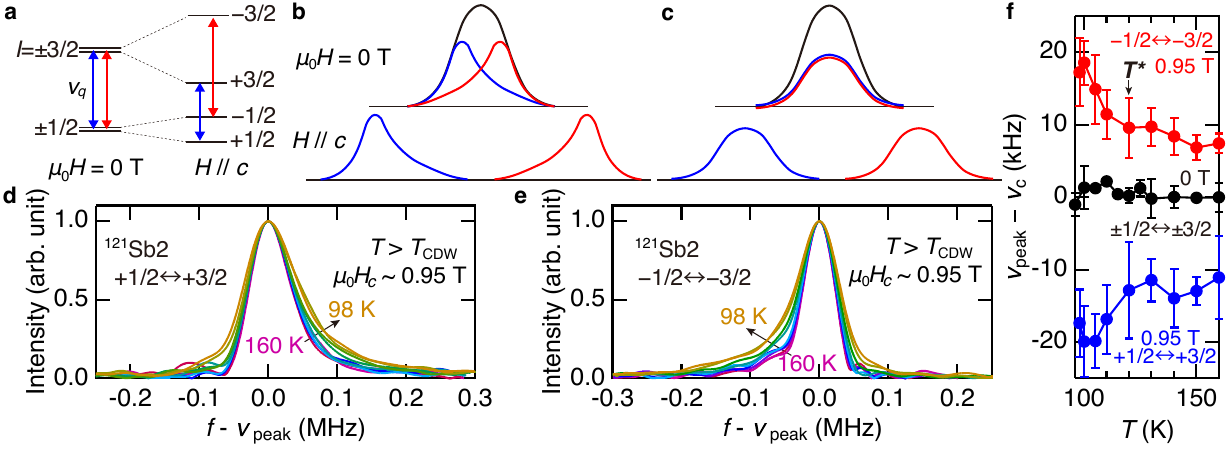}
	\caption{{\bf $^{121}$Sb2 NMR spectra and evidence of time-reversal symmetry breaking.} {\bf a}, Energy level diagram of $^{121}$Sb nuclear spin. In zero field, the electric quadrupole interaction splits the six-fold degenerate $I=5/2$ manifold into three doubly degenerate levels: $I=\pm 1/2$, $\pm 3/2$, and $\pm 5/2$. For simplicity, we focus on the $\pm 1/2 \leftrightarrow \pm 3/2$ transitions. An external magnetic field $\bm{H}||c$ further lifts the degeneracy by the Zeeman interaction, producing distinct resonances for the $+1/2 \leftrightarrow +3/2$ (blue) and $-1/2 \leftrightarrow -3/2$ (red) transitions. {\bf b}, {\bf c}, Schematic illustrations of linewidth broadening induced by magnetic (b) and quadrupole (c) interactions. The blue and red lines represent the spectra for $+1/2 \leftrightarrow +3/2$ and $-1/2 \leftrightarrow -3/2$ transitions, respectively. In the case of magnetic broadening, the hyperfine field induces frequency shifts of opposite sign for the two transitions, leading to antisymmetric lineshapes. In contrast, the quadrupole shift is independent of the spin sign, yielding symmetric lineshape for these transitions. {\bf d}, {\bf e}, $^{121}$Sb2 NMR spectra of the $+1/2 \leftrightarrow +3/2$ (d) and $-1/2 \leftrightarrow -3/2$ (e) transitions measured at $\mu_0 H_c \sim 0.95$\,T for $T > T_\mathrm{CDW}$. The normalized intensity is plotted as a function of $f - \nu_\mathrm{peak}$ to highlight the anomalous linewidth broadening, where $\nu_\mathrm{peak}$ is the peak frequency. The spectra show asymmetric broadening with temperature-dependent tails. The antisymmetric character of the lineshapes between the two transitions offer direct microscopic evidence for the presence of local magnetic fields. {\bf f}, Temperature dependence of the difference between the peak frequency $\nu_\mathrm{peak}$ and the spectral centroid $\nu_\mathrm{c}$ for the $- 1/2 \leftrightarrow - 3/2$ (red) and $+1/2 \leftrightarrow + 3/2$ (blue) transitions at 0.95\,T, and the $\pm 1/2 \leftrightarrow \pm 3/2$ transition at 0\,T (black). Reflecting the antisymmetric lineshapes, $\nu_\mathrm{peak}-\nu_\mathrm{c}$ increases for the $-1/2 \leftrightarrow -3/2$ transition and decreases for the $+1/2 \leftrightarrow +3/2$ transition below $T^* \approx 120$\,K. The error bars represent systematic uncertainties estimated by varying the intensity threshold used for the centroid analysis.
	\label{fig:NMR}
	}
\end{figure*}

\begin{figure*}
	\includegraphics[clip,width=15cm]{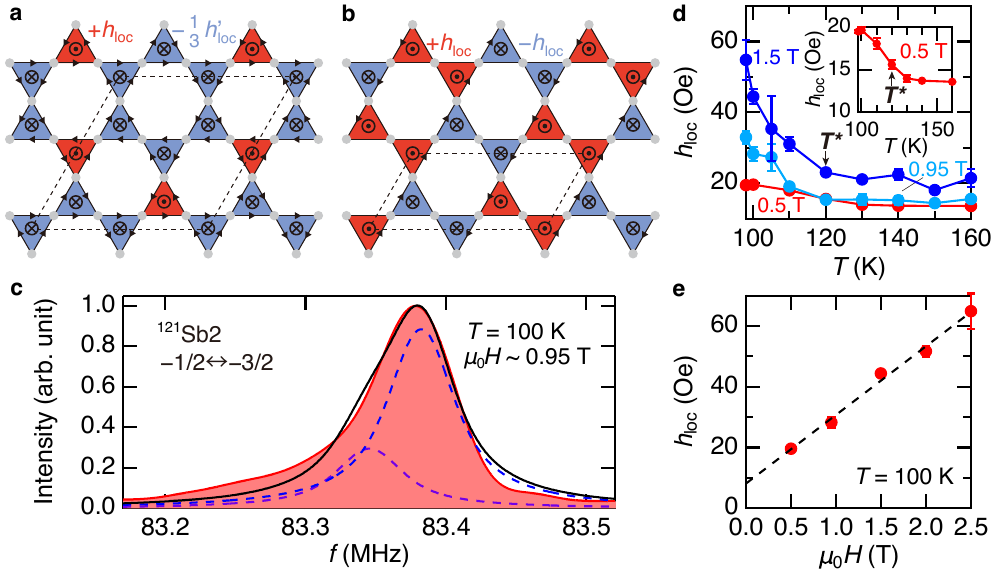}
	\caption{{\bf Loop current order and internal magnetic fields.} {\bf a}, Triple-$\bm{q}$ iCDW state. This configuration generates asymmetric local magnetic fields at Sb2 sites, $+h_\mathrm{loc}$ (red regions) and $-\frac{1}{3}h'_\mathrm{loc} (\approx -\frac{1}{3}h_\mathrm{loc})$ (blue regions) in a 1:3 ratio. The unit cell is depicted by the dashed line. {\bf b}, Single-$\bm{q}$ iCDW state. This configuration produces symmetric local fields at Sb2 sites, $+h_\mathrm{loc}$ (red regions) and $-h_\mathrm{loc}$ (blue regions) in a 1:1 ratio. {\bf c}, $^{121}$Sb2 NMR spectra measured at $\mu_0 H_c \sim 0.95$\,T and $T = 100$\,K. The experimental data (red filled area) are fitted well using a spectral model for the triple-$\bm{q}$ state, $I_\mathrm{total}(f, \Delta) = \frac{A}{3}V(f + \Delta) + AV(f - \frac{1}{3}\Delta)$ (see Methods). The fit (black solid line) is composed of two component peaks (blue and purple dashed lines). {\bf d}, Temperature dependence of the internal field $h_\mathrm{loc} (=\Delta/\gamma\mu_0)$ at $\mu_0H_c =$ 0.5, 0.95, and 1.5\,T. The data at 0.5\,T is enlarged in the inset. A sharp increase in $h_\mathrm{loc}$ occurs below $T^* \approx 120$\,K. {\bf e}, External magnetic field dependence of $h_\mathrm{loc}$ at $T=100$\,K. The internal field $h_\mathrm{loc}$ almost linearly increases with fields. The extrapolation to zero field yields a finite $h_\mathrm{loc}$ (dashed line), which provides compelling evidence for spontaneous time reversal symmetry breaking. The error bars represent systematic uncertainties in the fitting parameters estimated by varying the intensity threshold used for the fits.
	\label{fig:Loop_current}
	}
\end{figure*}

\clearpage

\renewcommand{\figurename}{Extended Data Fig. $\!\!$}
\setcounter{figure}{0}

\begin{figure*}
	\includegraphics[clip,width=6cm]{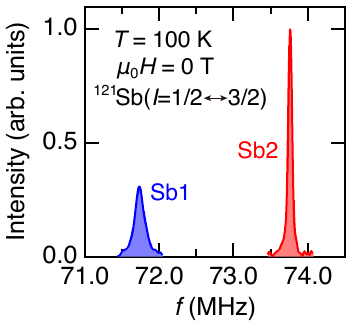}
	\caption{{\bf Zero-field NQR spectra of $^{121}$Sb ($I=5/2$) at 100\,K.} The spectra correspond to the $\pm 1/2 \leftrightarrow \pm 3/2$ transition in CsV$_3$Sb$_5$, which contains two inequivalent Sb sites. The relative intensities of the observed peaks reflect the atomic ratio of Sb1:Sb2 = 1:4.
	\label{fig:NQR_Sb12}
	}
\end{figure*}

\begin{figure*}
	\includegraphics[clip,width=6cm]{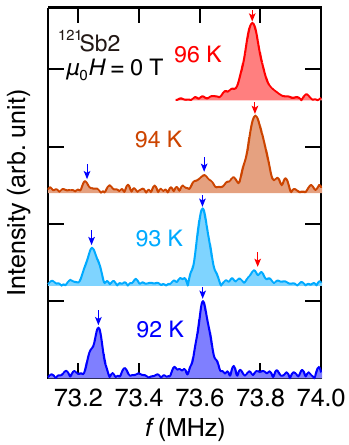}
	\caption{{\bf Evolution of the $^{121}$Sb2 NQR spectra near $T_\mathrm{CDW}$.} Below $T_\mathrm{CDW}$, the normal state peak (red arrows) is suppressed, while multiple peaks (blue arrows) appear due to the CDW modulation.
	\label{fig:CDW}
	}
\end{figure*}

\begin{figure*}
	\includegraphics[clip,width=16cm]{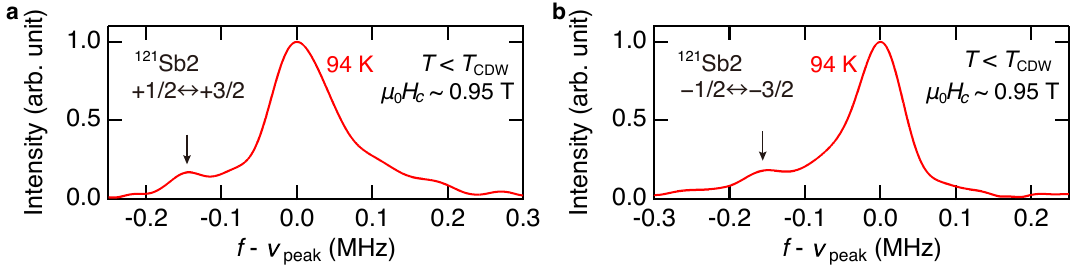}
	\caption{{\bf $^{121}$Sb2 NMR spectra below CDW transition temperature.} {\bf a}, {\bf b}, $^{121}$Sb2 NMR spectra of the $+1/2 \leftrightarrow +3/2$ (a) and $-1/2 \leftrightarrow -3/2$ (b) transitions at 94\,K. For both transitions, the CDW state peak (black arrows) appears at lower frequencies relative to the normal state peak.
	\label{fig:NMR_belowCDW}
	}
\end{figure*}

\begin{figure*}
	\includegraphics[clip,width=12cm]{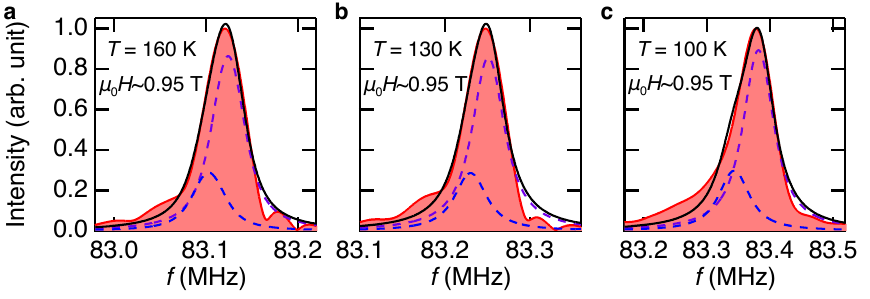}
	\caption{{\bf Temperature evolution of $^{121}$Sb2 NMR spectra measured at $\mu_0 H_c \sim 0.95$\,T.} {\bf a-c}, $^{121}$Sb2 NMR spectra measured at $T=160$\,K (a), 130\,K (b), and 100\,K (c). The experimental data (red) are fitted using the spectral model $I_\mathrm{total}(f, \Delta) = \frac{A}{3}V(f + \Delta) + AV(f - \frac{1}{3}\Delta)$ corresponding to the triple-$\bm{q}$ loop current state (see Methods). The black solid lines represent the total fit, comprising two component peaks (blue and purple dashed lines).
	\label{fig:Temp_fit}
	}
\end{figure*}

\begin{figure*}
	\includegraphics[clip,width=12cm]{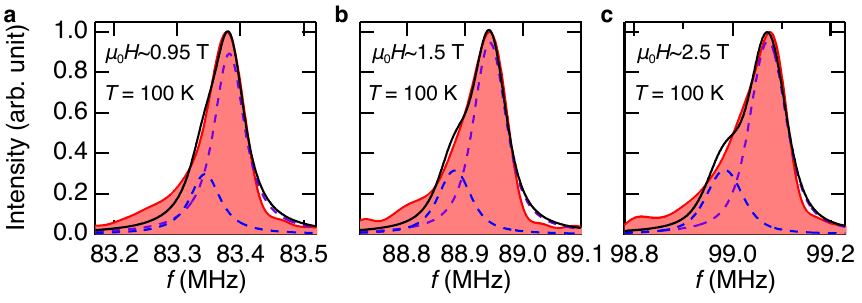}
	\caption{{\bf Field evolution of $^{121}$Sb2 NMR spectra at 100\,K.} {\bf a-c}, $^{121}$Sb2 NMR spectra at $\mu_0H \ sim 0.95$\,T (a), 1.5\,T (b), and 2.5\,T (c). The experimental data (red) are fitted using the model $I_\mathrm{total}(f, \Delta) = \frac{A}{3}V(f + \Delta) + AV(f - \frac{1}{3}\Delta)$ corresponding to the triple-$\bm{q}$ loop current state (see Methods). The black solid lines represent the total fit, comprising two component peaks (blue and purple dashed lines).
	\label{fig:Field_fit}
	}
\end{figure*}

\begin{figure*}
	\includegraphics[clip,width=12cm]{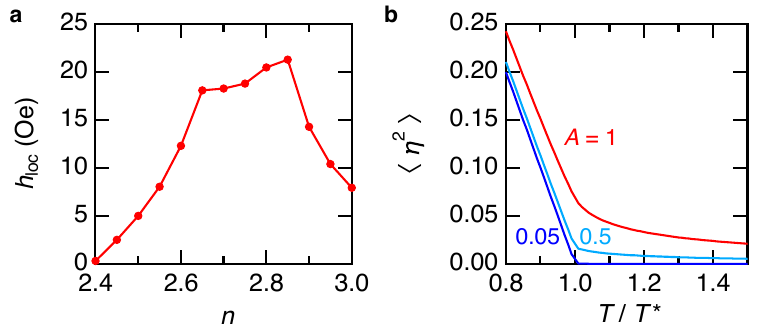}
	\caption{{\bf Numerical calculation of emergent local fields and linewidth broadening.} {\bf a}, Numerical calculation of the filling $n$ dependence of internal field $h_\mathrm{loc}$ in the triple-$\bm{q}$ loop current state in zero field. For CsV$_3$Sb$_5$, where $n \sim 3$, the calculated internal fields of approximately 10\,Oe is consistent with the experimental results (see Fig.\,4). {\bf b}, Numerical calculation of mean square fluctuation $\langle \eta^2 \rangle$ incorporating thermal fluctuations. For strong thermal fluctuations ($A \sim 1$), finite $\langle \eta^2 \rangle$ above $T^*$ leads to finite internal fields in this temperature range (see Methods for details).
	\label{fig:calculation}
	}
\end{figure*}

\begin{figure*}
	\includegraphics[clip,width=6cm]{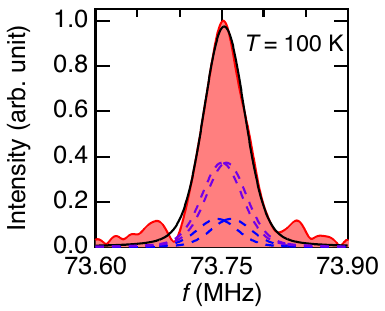}
	\caption{{\bf Zero-field NQR spectra at 100\,K.} In zero field, the $+1/2 \leftrightarrow +3/2$ and $-1/2 \leftrightarrow -3/2$ transitions overlaps, leading to $I_\mathrm{total}(f, \Delta) = \frac{A}{3}V(f + \Delta) + AV(f - \frac{1}{3}\Delta) + \frac{A}{3}V(f - \Delta) + AV(f + \frac{1}{3}\Delta)$ corresponding to the triple-$\bm{q}$ loop current state (see Methods). The experimental data (red) are well reproduced by this spectral form using the extrapolated zero-field value $h_\mathrm{loc} = 8$\,Oe obtained from the analysis on the field dependence (Fig.\,4e). The black solid lines represent the total fit, comprising four component peaks (blue and purple dashed lines).
	\label{fig:NQR_fixed_hloc}
	}
\end{figure*}

\begin{figure*}
	\includegraphics[clip,width=16cm]{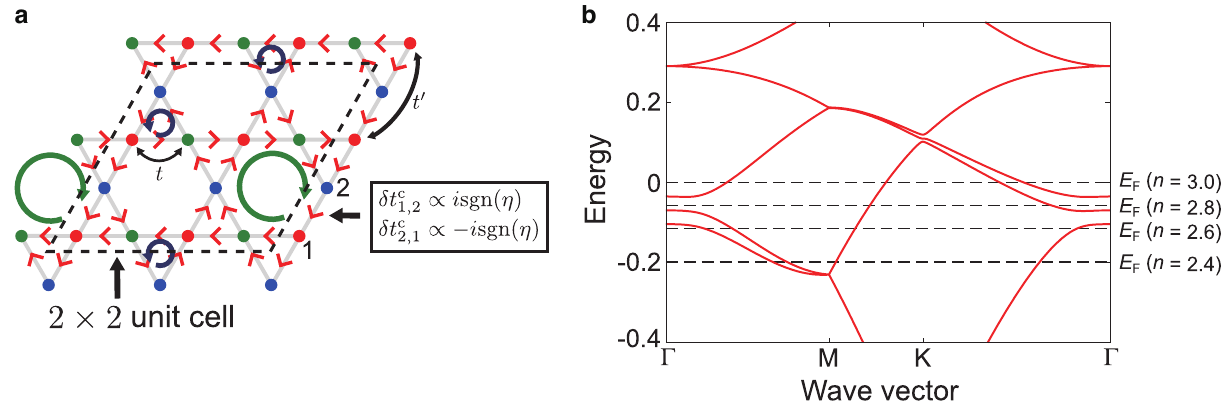}
	\caption{{\bf Theoretical model for numerical calculations of emergent local fields.} {\bf a}, Kagome lattice tight-binding model with a $2 \times 2$ iCDW order. The iCDW order is given by a pure imaginary modulation of the nearest neighbor hopping between V sites $(i,j)$; $\delta t_{i,j} = i\eta$ or $-i\eta$. The unit cell is depicted by the dashed line. {\bf b}, Band structure of the iCDW state calculated for $\eta = -0.01$\,eV. Horizontal dashed lines denote the Fermi energy for different filling $n$. For CsV$_3$Sb$_5$, the relevant filling is $n \sim 3$.
	\label{fig:band}
	}
\end{figure*}

\end{document}



\title{Supplementary Information for Microscopic evidence for imaginary charge density wave in a kagome metal}


\author{S. Suetsugu}
\affiliation{Department of Applied Physics, The University of Tokyo, Tokyo 113-8656, Japan}
\affiliation{Department of Physics, Kyoto University, Kyoto 606-8502, Japan}
\author{F. Hori}
\affiliation{Department of Physics, Kyoto University, Kyoto 606-8502, Japan}
\affiliation{Department of Physics, Graduate School of Science, Tohoku University, 6-3 Aramaki-Aoba, Aoba-ku, Sendai, Miyagi 980-8578, Japan}
\author{M. Shibata}
\affiliation{Department of Physics, Kyoto University, Kyoto 606-8502, Japan}
\author{S. Kitagawa}
\affiliation{Department of Physics, Kyoto University, Kyoto 606-8502, Japan}
\author{K. Ishida}
\affiliation{Department of Physics, Kyoto University, Kyoto 606-8502, Japan}
\author{T. Asaba }
\affiliation{Department of Physics, Kyoto University, Kyoto 606-8502, Japan}
\affiliation{Department of Physics, University of Virginia, Charlottesville, VA. 22904, USA}
\author{S. Nakazawa}
\affiliation{Department of Physics, Nagoya University, Furo-cho, Nagoya 464-8602, Japan}
\author{Q. Li}
\affiliation{National Laboratory of Solid State Microstructures and Department of Physics, Nanjing University, Nanjing, China}
\author{H. -H. Wen}
\affiliation{National Laboratory of Solid State Microstructures and Department of Physics, Nanjing University, Nanjing, China}
\author{T. Shibauchi}
\affiliation{Department of Advanced Materials Science, University of Tokyo, Chiba, Japan}
\author{H. Kontani}
\affiliation{Department of Physics, Nagoya University, Furo-cho, Nagoya 464-8602, Japan}
\author{Y. Matsuda}
\affiliation{Department of Physics, Kyoto University, Kyoto 606-8502, Japan}
\affiliation{Los Alamos National Laboratory, Los Alamos, New Mexico 87545, USA}
\date{\today}


\maketitle
%

\section*{NQR/NMR spectra of $^\mathbf{121}\mathbf{Sb1}$}
\begin{figure*}
	\includegraphics[clip,width=10cm]{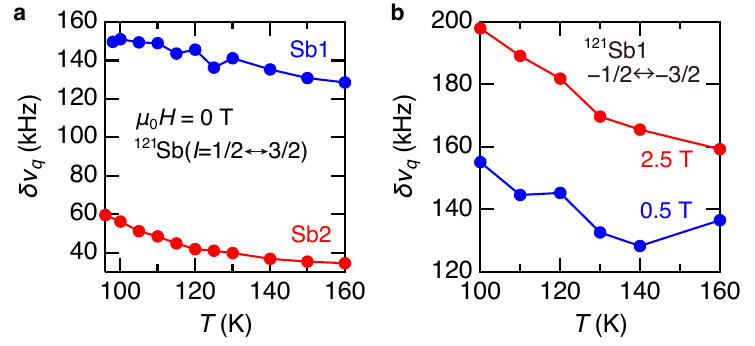}
	\caption{Spectral linewidth $\delta\nu_q$ at the $^{121}$Sb1 site. {\bf a}, Temperature dependence of the linewidth $\delta\nu_q$ for the $\pm 1/2 \leftrightarrow \pm 3/2$ transition at the Sb1 site (blue). For comparison, we also plot the data for the Sb2 site. The linewidth at the Sb1 site is significantly larger than that at the Sb2 site and increases gradually upon cooling. {\bf b}, Temperature dependence of $\delta\nu_q$ for the $- 1/2 \leftrightarrow - 3/2$ transition at the Sb1 site measured at $\mu_0 H = 0.5$\,T (blue) and 2.5\,T (red). The linewidth exhibits an enhancement below $T^* \sim 120$\,K.
	\label{fig:Sb1_width}
	}
\end{figure*}

We performed NQR and NMR measurements at the Sb1 site. Compared with the Sb2 site, the Sb1 spectra are significantly broader and weaker in intensity (Extended Data Fig.\,1), which makes a detailed quantitative analysis more challenging. As shown in Fig.\,\ref{fig:Sb1_width}a, the NQR linewidth $\delta\nu_q$ of the Sb1 site is much larger than that of the Sb2 site over the entire temperature range. Although the linewidth enhancement upon cooling is less pronounced than that observed at the Sb2 site, the Sb1 linewidth gradually increases with decreasing temperature. In magnetic fields, the NMR signal from the Sb1 site becomes even weaker, preventing a detailed analysis of the temperature evolution of the lineshape comparable to that for the Sb2 site. Therefore, for the field-dependent measurements, we focus on the temperature dependence of the linewidth (Fig.\,\ref{fig:Sb1_width}b). The linewidth exhibits an enhancement below $T^* \sim 120$\,K, which is consistent with the emergence of an internal magnetic field at the Sb1 site.

\begin{figure*}
	\includegraphics[clip,width=16cm]{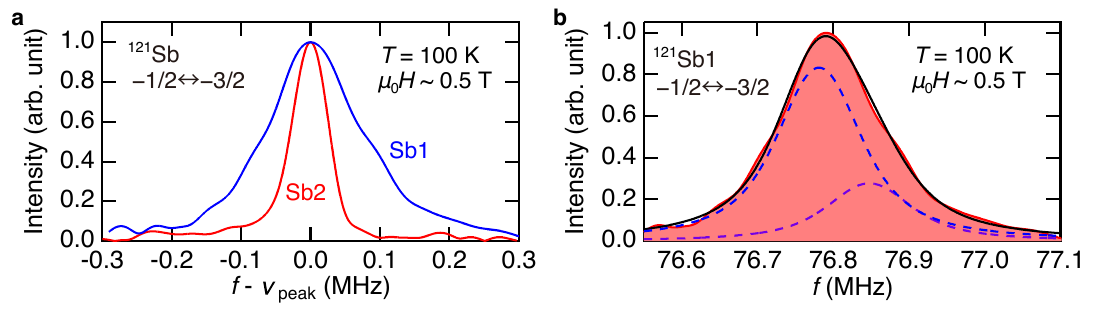}
	\caption{$^{121}$Sb1 NMR spectra. {\bf a}, Comparison of $^{121}$Sb1 and $^{121}$Sb2 NMR spectra measured at $\mu_0 H \sim 0.5$\,T and $T = 100$\,K. The normalized intensity is plotted as a function of $f - \nu_q$ to highlight the asymmetric broadening. While the Sb2 spectrum display a tail extending toward lower frequencies, the Sb1 spectrum shows a tail extending toward higher frequencies. This opposite asymmetry is consistent with the triple-$\bm{q}$ loop current ordered state. {\bf b}, $^{121}$Sb1 NMR spectrum measured at $\mu_0 H_c \sim 0.5$\,T and $T = 100$\,K. The experimental data (red filled area) are well fitted by a spectral model for the triple-$\bm{q}$ state, $I_\mathrm{total}(f, \Delta) = \frac{A}{3}V(f + \Delta) + AV(f - \frac{1}{3}\Delta)$ (see Methods). The fit (black solid line) is composed of two component peaks (blue and purple dashed lines). The best fit is obtained for $h^\mathrm{Sb1}_\mathrm{loc} \sim -50$\,Oe.
	\label{fig:Sb1_spectra}
	}
\end{figure*}

To further examine the origin of this internal field, we compare the NMR spectra of the $-1/2 \leftrightarrow -3/2$ transitions of the Sb1 and Sb2 sites measured at $T = 100$\,K and $\mu_0 H = 0.5$\,T (Fig.\,\ref{fig:Sb1_spectra}a). While the Sb2 spectrum exhibits a tail extending toward lower frequencies, the Sb1 spectrum shows a tail on the high-frequency side. In the triple-$\bm{q}$ loop current ordered state, the Sb1 sites experience hyperfine fields of $\Delta H = h^\mathrm{Sb1}_\mathrm{loc}$ and $-\frac{1}{3}h^{'\mathrm{Sb1}}_\mathrm{loc}$ in a 1:3 ratio. While this two-component structure of the local field is similar to that at the Sb2 site, the corresponding field components have opposite signs between the Sb1 and Sb2 sites. As a result, the spectral asymmetry, and hence the direction of the spectral tail, is reversed between the two sites. Therefore, the observed contrast between the Sb1 and Sb2 spectra is consistent with the triple-$\bm{q}$ scenario.

To quantitatively estimate the local field at the Sb1 site, we fitted the Sb1 spectra using a model incorporating the above two-component field structure. As shown in Fig.\,\ref{fig:Sb1_spectra}b, this model provides excellent agreement with the experimental data, with the best fit yielding $h^\mathrm{Sb1}_\mathrm{loc} \sim -50$\,Oe. The magnitude of this value is larger than the local field $h_\mathrm{loc} \sim 20$\,Oe extracted for the Sb2 site (Fig.\,4e), consistent with the distinct crystallographic environments of the two Sb sites. Indeed, numerical calculations of internal fields in zero field predict that the local field at the Sb1 site is several times larger than that at the Sb2 site (Fig.\,S3). Taken together, these results provide further support for the triple-$\bm{q}$ loop current ordered state.

\begin{figure*}
	\includegraphics[clip,width=6cm]{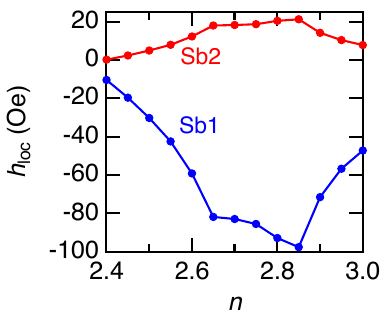}
	\caption{Numerical calculation of the filling $n$ dependence of internal field $h_\mathrm{loc}$ in the triple-$\bm{q}$ state in zero field. The filling corresponding to CsV$_3$Sb$_5$ is $n \sim 3$. The calculated internal field at the Sb1 site is several times larger than that at the Sb2 site.
	\label{fig:Sb1_calculation}
	}
\end{figure*}

\section*{Effect of domain structure of loop current ordered state}
The triple-$\bm{q}$ loop current ordered state is expected to host two types of domains with opposite circulating current directions and associated local fields. NMR/NQR spectra originating from these two domains exhibit asymmetric tails extending to opposite sides of the resonance frequency. If the two domains were present in equal proportions, their superposition would result in a symmetric lineshape, even in the presence of internal local fields.

However, the triple-$\bm{q}$ state has a finite uniform orbital magnetization \cite{Tazai-Morb}. Consequently, under an applied magnetic field, one of the two domains is energetically favored, which can lead to partial or complete domain alignment. The asymmetric spectral lineshapes observed in our NMR experiments thus suggest a field-induced alignment of the loop current domains.

\section*{Effect of a finite asymmetry parameter}
Previous studies have reported that the electric field gradient (EFG) asymmetry parameter $\eta$ at the Sb2 site vanishes above the CDW transition temperature \cite{mu2021s}. Accordingly, we focus on the analysis for $\eta = 0$ in the main text. In this section, we extend the discussion to the case of a finite asymmetry parameter $\eta$.

For $\eta \neq 0$ and in the presence of an external magnetic field $\bm{H}||c$, the nuclear spin Hamiltonian at the $^{121}$Sb2 site ($I=5/2$) is written as
\begin{align}
\mathcal{H} &= \mathcal{H}_Q + \mathcal{H}_Z \label{eqn:Hamiltonian} \\
\mathcal{H}_Q &= \frac{h\nu_{q}}{6}\left(
3I_z^2 - I^2 + \frac{\eta}{2}\left(I_+^2 + I_-^2\right)
\right) \\
\mathcal{H}_Z &= - \gamma \hbar \mu_0 (H + \Delta H) I_z ,
\end{align}
where $\mathcal{H}_Q$ and $\mathcal{H}_Z$ denote the electric quadrupole and Zeeman interaction terms, respectively. Here $\nu_q$ is the quadrupole coupling constant, and $\Delta H$ represents the hyperfine field at the nuclear site. For finite $\eta$, $I_z$ is no longer a good quantum number owing to the mixing of different spin components. Nevertheless, in the regime of small $\eta$, the transitions can still be unambiguously labeled by their dominant $I_z$ components. Hereafter, transitions with predominantly $+1/2 \leftrightarrow +3/2$ and $-1/2 \leftrightarrow -3/2$ character are simply referred to as the $+1/2 \leftrightarrow +3/2$ and $-1/2 \leftrightarrow -3/2$ transitions, respectively.

\begin{figure*}
	\includegraphics[clip,width=14cm]{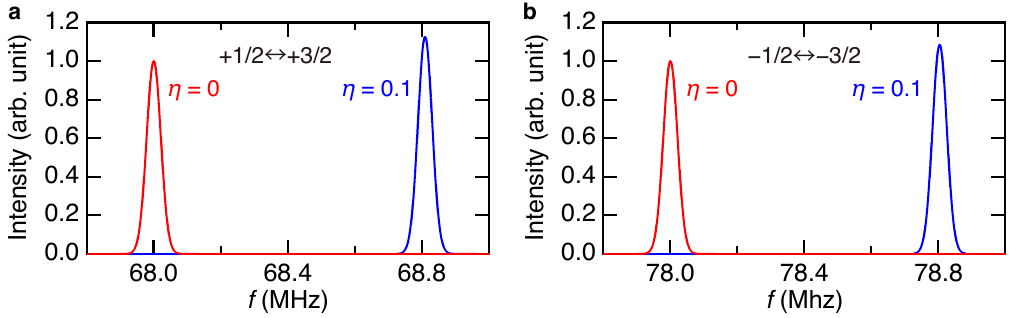}
	\caption{Calculated NMR spectra for $\eta = 0$ (red) and $\eta = 0.1$ (blue). For both the transitions with predominantly $+1/2 \leftrightarrow +3/2$ character (a) and the $-1/2 \leftrightarrow -3/2$ character (b), a finite asymmetry parameter $\eta$ shifts the resonant frequencies toward higher frequencies. The numerical calculations are performed using $\nu_q = 73$\,MHz and $\nu_L = \gamma \mu_0 (H + \Delta H) / 2\pi = 5$\,MHz.
	\label{fig:eta_compare}
	}
\end{figure*}

Figure~\ref{fig:eta_compare} shows NMR spectra obtained from numerical calculations based on the Hamiltonian in Eq.\,(\ref{eqn:Hamiltonian}), for $\nu_q = 73$\,MHz and $\nu_L = \gamma \mu_0 (H + \Delta H) / 2\pi = 5$\,MHz, comparing the cases of $\eta = 0$ (red) and $\eta = 0.1$ (blue). For both the $+1/2 \leftrightarrow +3/2$ and the $-1/2 \leftrightarrow -3/2$ transitions, a finite $\eta$ shifts the resonance frequencies toward higher frequencies. However, a uniform distribution of a finite asymmetry paramter cannot account for the asymmetric lineshape observed in our experiments (Figs.\,3d and e).

\begin{figure*}
	\includegraphics[clip,width=14cm]{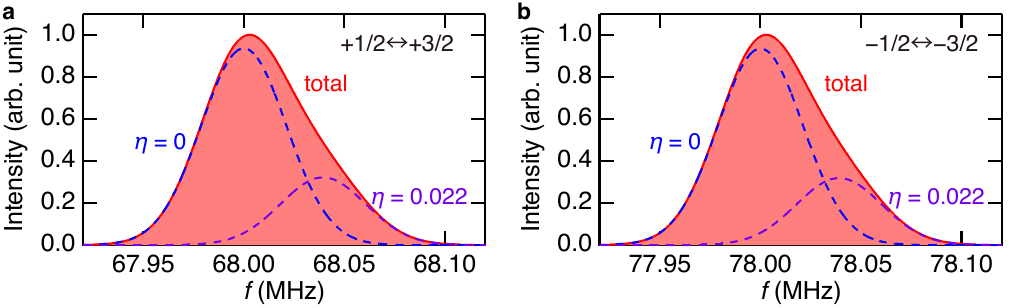}
	\caption{Calculated NMR spectra assuming an inhomogeneous distribution of the EFG asymmetry parameter $\eta$. The spectra are constructed as a weighted sum of two components (red filled area) corresponding to $\eta = 0$ and $\eta = 0.022$ in a 3:1 ratio (blue and purple dashed lines), calculated with $\nu_q = 73$\,MHz and $\nu_L = 5$\,MHz. For both the transitions with predominantly $+1/2 \leftrightarrow +3/2$ (a) and the $-1/2 \leftrightarrow -3/2$ (b) character, the spectra exhibit tails extending toward higher frequencies.
	\label{fig:eta_asym}
	}
\end{figure*}

To reproduce the asymmetric lineshape, it is necessary to assume an inhomogeneous distribution of sites with different values of $\eta$. As an illustrative example, Fig.\,\ref{fig:eta_asym} shows a weighted sum of two components corresponding to $\eta = 0$ and $\eta = 0.022$ in a 3:1 ratio, calculated with $\nu_q = 73$\,MHz and $\nu_L = 5$\,MHz. For the $+1/2 \leftrightarrow +3/2$ transition, this model yields a spectrum exhibiting a pronounced tail extending toward higher frequencies. However, the same distribution of $\eta$ values inevitably produces a spectrum with a high-frequency tail also for the $-1/2 \leftrightarrow -3/2$ transition. This is because a finite $\eta$ shifts the resonant frequencies of both transitions in the same direction, leading to tails on the same side for the $+1/2 \leftrightarrow +3/2$ and $-1/2 \leftrightarrow -3/2$ spectra. Therefore, a finite EFG asymmetry parameter cannot account for the experimentally observed antisymmetric lineshapes between the $+1/2 \leftrightarrow +3/2$ and $-1/2 \leftrightarrow -3/2$ transitions. These results further corroborate the magnetic origin of the anomalous linewidth broadening.

\clearpage

\bibliography{ref_supp.bib}